# FEniCS implementation of the Virtual Fields Method (VFM) for nonhomogeneous hyperelastic identification

Jianwei Deng[1,2], Xu Guo[1,2,3], Yue Mei[1,2,3,4 *], Stephane Avril[5*]

[1]State Key Laboratory of Structural Analysis for Industrial Equipment, Department of Engineering Mechanics, Dalian University of Technology, Dalian 116023, P.R. China

[2]International Research Center for Computational Mechanics, Dalian University of Technology, Dalian 116023, P.R. China

[3]Ningbo Institute of Dalian University of Technology, Ningbo, China

[4]DUT-BSU Joint Institute, Dalian University of Technology, 116023, P.R. China

[5]Mines Saint-Étienne, Univ Lyon, Univ Jean Monnet, INSERM, U 1059 Sainbiose, 42023, Saint-Étienne, France

Corresponding author*: meiyue@dlut.edu.cn (Yue Mei) and avril@emse.fr (Stephane Avril)

## Abstract

It is of great significance to identify the nonhomogeneous distribution of material properties in human tissues for different clinical and medical applications. This leads to the requirement of solving an inverse problem in elasticity. The virtual fields method (VFM) is a rather recent inverse method with remarkable computational efficiency compared with the optimization-based methods. In this study, we aim to identify nonhomogeneous hyperelastic material properties using the VFM. We propose two novel algorithms, RE-VFM and NO-VFM. In RE-VFM, the solid is partitioned in different regions and the elastic properties of each region are determined. In NO-VFM, the distribution of elastic properties is completely reconstructed through the inverse problem without partitioning the solid. As the VFM requires to use virtual fields, we proposed an efficient way to construct them and implemented the approach in the FEniCS package. We validated the proposed methods on several examples, including a bilayer structure, a lamina cribosa (LC) model and a cube model embedded with a spherical inclusion. The numerical examples illustrate the feasibility of both RE-VFM and NO-VFM. Notably, the spatial variations of the Young's modulus distribution can be recovered accurately within only 5 iterations. The obtained results reveal the potential of the proposed methods for future clinical applications such as estimating the risk of vision loss related to glaucoma and detecting tumors.

# Introduction

Extensive studies have shown that distribution of material properties in human tissues varies with age or disease [1]–[3]. Therefore, the reconstruction of material parameters can provide valuable information for clinical diagnosis from the perspective of mechanics. To obtain the mechanical property of soft tissues, we should solve inverse problems in elasticity. Utilizing deformation fields provided by full-field measurement techniques such as Ultrasound [4], Magnetic Resonance Imaging (MRI) [5] and Optical Coherence Tomography (OCT) [6] has become more and more commonplace.

The Virtual Fields Method is a typical inverse method based on the principle of virtual power. Compared with the prevalent optimization-based inverse method [7]–[9], the VFM is often more computationally efficient. Thus, the VFM has become more and more extensively used to solve inverse problems [2], [10]–[12].

For nonhomogeneous solids, the domain of interest is usually partitioned into a finite number of homogeneous regions. It is assumed that the boundaries of these regions are known when the VFM is applied. Accordingly, only the biomechanical property of each region needs to be determined [13], [14]. However, if the boundary of each homogeneous region is unknown, the inverse problem becomes highly challenging to solve due to the large number of unknown elastic parameters. To address this issue, a novel scheme considering virtual work balance between neighborhood elements was recently proposed [15]. However, this previous study focused on a simple geometry. Besides, the choice of virtual fields was a key issue because virtual fields influence the accuracy of the estimation. For models with complex geometries, the construction of optimal virtual fields is usually non-trivial. Moreover, for nonlinear elastic solids with multiple material parameters, more virtual fields are required to avoid ill-conditioned systems, which increases the difficulty of constructing the virtual fields.

There are various suitable constitutive models to describe the mechanical behavior for different soft tissues, such as anisotropic hyperelastic models [16] [17] and viscoelastic models [18] [19]. In this paper, for the purpose of simplicity, we adopt the isotropic Neo-Hookean constitutive model for testing the proposed algorithms.

In this study, we propose two VFM-based inverse methods to identify the nonhomogeneous distribution of hyperelastic solids: RE-VFM and NO-VFM methods. "RE" is short for "region", namely the material parameters for every region are considered as unknowns. In the RE-VFM method, the boundary of each region is assumed to be known as a *priori*. To this end, regional material parameters are required to be estimated. And "NO" is short for "nodal", namely, the material parameters for every node are considered as unknowns. In the NO-VFM method, the boundary of each region is unknown, thus, all nodal material properties are demanded to estimate. In this paper, we also propose an efficient way to construct virtual fields for the identification problem, which is implemented in the open-source FEniCS platform [20] using Python. We eventually make the proof of concept of the proposed methods by several numerical examples.

The paper is organized as follows: in the **Method** section, we elaborate the mathematical aspects of the proposed VFM inverse methods; in the **FEniCS implementation** section, we discuss the details of the implementation and of the algorithms; in the **Results** section, numerical tests are performed and results are shown; the **Conclusions** section closes the paper.

## Method

### Inverse problem

Suppose we have a hyperelastic solid $\beta$ or part of it, in the reference configuration $\kappa_R(\beta)$, under the given tractions **t** on the traction boundary $\Gamma_t$ of $\beta$. We denote a material parameter vector $P = \{P_1, P_2, ..., P_q, ..., P_N\}^T$ of $\beta$, which drives the nonlinear elastic response of $\beta$ under the given tractions and boundary conditions. $\beta$ undergoes a motion which can be described by a mapping $\chi_P$ from the reference configuration $\kappa_R(\beta)$ to the current configuration $\kappa_C(\beta)$,

$$\begin{aligned}
\mathbf{x} &= \chi_P(\mathbf{X}, t) = \mathbf{X} + \mathbf{u}(P, \mathbf{X}, t) \\
\mathbf{F} &= \frac{\partial \chi_P(\mathbf{X}, t)}{\partial \mathbf{X}} = \mathbf{I} + \frac{\partial \mathbf{u}(P, \mathbf{X}, t)}{\partial \mathbf{X}} \\
\mathbf{C} &= \mathbf{F}^T \mathbf{F}
\end{aligned} \qquad (1)$$

where **x** and **X** are position vectors in the current configuration at time $t$ and reference configuration, respectively, **u** is the displacement field, **F** is the deformation gradient tensor and **C** is the right Cauchy-Green tensor.

We assume that an experimental measurement $\mathbf{u}_{meas}$ of the actual displacement field $\mathbf{u}(\mathbf{X}, t)$ at time $t$ is obtained by a full-field measurement technique, possibly with some measurement noise. We can also derive the corresponding $\mathbf{F}_{meas}$ and $\mathbf{C}_{meas}$ by $\mathbf{u}_{meas}$. The inverse problem is defined as: finding an estimated parameter vector $\tilde{P}$ such that the gap between $\mathbf{u}_{meas}$ and $\mathbf{u}(\tilde{P}, \mathbf{X}, t)$ is minimized.

### Definition of the intermediate configuration

For solving the inverse problem, we define an intermediate configuration $\kappa_{p^o}(\beta)$ for the solid $\beta$, where $P^o$ corresponds to an initial estimation of the unknown material parameters. The position vector in the intermediate configuration can be denoted as $\mathbf{x}^o = \chi_{P^o}(\mathbf{X},t)$, and the corresponding displacement $\mathbf{u}^o(\mathbf{X},t) = \mathbf{x}^o - \mathbf{X}$.

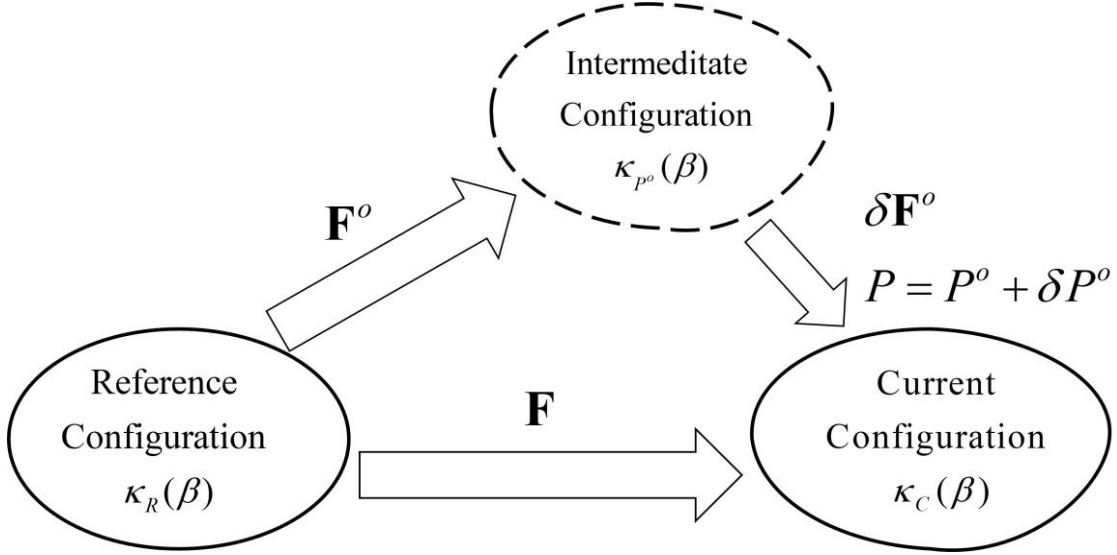

**Figure 1** Illustration of the intermediate configuration.

In this intermediate configuration, we suppose a small displacement $\delta\mathbf{x}^o$ superimposed upon $\mathbf{u}^o$, yielding the current position $\mathbf{x}$ at time $t$ for the material parameters $P$ at equilibrium. A small variation of displacements and parameters around the intermediate configuration is considered.

$$\begin{aligned} \mathbf{x} &= \mathbf{x}^o + \delta\mathbf{x}^o \quad (\|\delta\mathbf{x}^o\| \ll \|\mathbf{u}^o\|) \\ P &= P^o + \delta P^o \quad (\|\delta P^o\| \ll \|P^o\|) \end{aligned} \qquad (2)$$

The deformation gradient with the mapping from the reference to the intermediate configuration $\mathbf{F}^o$ and the deformation gradient with the mapping from the intermediate configuration to the current configuration $\delta\mathbf{F}^o$ are given as:

$$\begin{aligned} \mathbf{F}^o &= \frac{\partial \chi_{P^o}(\mathbf{X},t)}{\partial \mathbf{X}} \\ \delta\mathbf{F}^o &= \frac{\partial \delta\mathbf{x}^o}{\partial \mathbf{x}^o} \end{aligned} \qquad (3)$$

The relationship between the deformation gradients and the right Cauchy-Green tensor between the successive motion can be written as:

$$\begin{aligned}&\mathbf{F}=(\mathbf{I}+\delta\mathbf{F}^o)\mathbf{F}^o=\mathbf{F}^o+\delta\mathbf{F}^o\mathbf{F}^o\\&\mathbf{C}=\mathbf{C}^o+\delta\mathbf{C}^o\\&\mathbf{C}^o=(\mathbf{F}^o)^T\mathbf{F}^o\end{aligned} \quad (4)$$

**Equilibrium equations**

In the absence of the body force, the quasi-static equilibrium equation of forces responsible for deformation $\mathbf{F}$ may be written in the reference configuration as:

$$\begin{aligned}&\nabla\cdot(\mathbf{FS})=0 \text{ on } \beta\\&(\mathbf{FS})\cdot\mathbf{n}=\mathbf{t} \text{ on } \Gamma_t\end{aligned} \quad (5)$$

where $\nabla\cdot$ denotes the divergence operator in the reference configuration, $\mathbf{S}$ is the 2$^{nd}$ Piola-Kirchoff stress tensor, $\mathbf{n}$ is the unit exterior normal vector of $\Gamma_t$ in the reference configuration, and $\mathbf{t}$ is the applied traction in the reference configuration.

Similarly, quasi-static equilibrium equation of forces responsible for deformation $\mathbf{F}^o$ may be written in the reference configuration as:

$$\begin{aligned}&\nabla\cdot(\mathbf{F}^o\mathbf{S}^o)=0 \text{ on } \beta\\&(\mathbf{F}^o\mathbf{S}^o)\cdot\mathbf{n}=\mathbf{t} \text{ on } \Gamma_t\end{aligned} \quad (6)$$

The 2$^{nd}$ Piola-Kirchoff stress $\mathbf{S}$ can be related to $\mathbf{S}^o$ by a first-order Taylor expansion. Substituting $P=P^o+\delta P^o$ and $\mathbf{C}=\mathbf{C}^o+\delta\mathbf{C}^o$, $\mathbf{S}$ can be expressed by:

$$\mathbf{S}\approx\mathbf{S}^o+\delta\mathbf{S}^o\approx\mathbf{S}^o+\frac{\partial\mathbf{S}}{\partial\mathbf{C}}:\delta\mathbf{C}^o+\Sigma_{q=1}^{N}\delta P_q^o\mathbf{S}^o_{,P_q} \quad (7)$$

where $\mathbf{S}^o_{,P_q}=\frac{\partial\mathbf{S}^o}{\partial P_q}, \delta\mathbf{C}^o=2\delta\mathbf{E}^o\simeq 2\mathbf{F}^{oT}\delta\boldsymbol{\varepsilon}^o\mathbf{F}^o$, $\delta\mathbf{E}^o=\mathbf{E}-\mathbf{E}^o=\frac{1}{2}(\mathbf{C}-\mathbf{I})-\frac{1}{2}(\mathbf{C}^o-\mathbf{I})$.

$\mathbf{E}$ and $\mathbf{E}^o$ are the Green strain tensor in the current and intermediate configuration, respectively. And $\delta\boldsymbol{\varepsilon}^o$ is the infinitesimal strain induced by an infinitesimal variation of material properties. When $\delta\mathbf{F}^o$ is small, $\delta\boldsymbol{\varepsilon}^o$ can be written as

$$\delta \boldsymbol{\varepsilon}^o = \frac{1}{2}(\delta \mathbf{F}^o + \delta \mathbf{F}^{oT}) \tag{8}$$

Yielding,

$$\delta \mathbf{S}^o = \mathbf{K}^o : \delta \mathbf{E}^o + \Sigma_{q=1}^{N} \delta P_q^o \mathbf{S}^o_{,P_q}$$

$$\mathbf{K}^o = 2\frac{\partial \mathbf{S}^o}{\partial \mathbf{C}^o} \tag{9}$$

Next, we substitute $\mathbf{S} = \mathbf{S}^o + \delta \mathbf{S}^o$ and $\mathbf{F} = \mathbf{F}^o + \delta \mathbf{F}^o \mathbf{F}^o$ into Eq. (5). Considering the equilibrium of $\mathbf{S}^o$ in Eq. (6), and neglecting the second-order small quantity $\delta \mathbf{F}^o \delta \mathbf{S}^o$ term, we finally have

$$\nabla.(\mathbf{F}^o \delta \mathbf{S}^o) = -\nabla.(\mathbf{F}^o \delta \mathbf{F}^o \mathbf{S}^o) \text{ on } \beta$$

$$(\mathbf{F}^o \mathbf{S}^o).\mathbf{n} = -(\mathbf{F}^o \delta \mathbf{F}^o \mathbf{S}^o).\mathbf{n} \text{ on } \Gamma_t \tag{10}$$

**Hyperelastic constitutive behavior**

In this part, although our approach can be generalized to any hyperelastic constitutive law, we introduce a typical hyperelastic constitutive law, the compressible Neo-Hookean model with material parameters vector written as $P = \{E, v\}^T$ where $E$ is the Young's modulus and $v$ is the Poisson's ratio. The strain energy density function $\Phi$ and the 2$^{nd}$ Piola-Kirchoff stress tensor $\mathbf{S}$ of a Neo-Hookean model can be written as [21]:

$$\Phi = \frac{1}{2}[\mu(I_1 - 3) - 2\mu \ln(J) + \lambda(\ln J)^2] \tag{11}$$

$$\mathbf{S} = 2\frac{\partial \Phi}{\partial \mathbf{C}} = \lambda(\ln J)\mathbf{C}^{-1} + \mu(\mathbf{I} - \mathbf{C}^{-1}) \tag{12}$$

where $I_1 = \text{trace}(\mathbf{C}), J = \det(\mathbf{F})$, $\mathbf{I}$ is the 2-order identity tensor. $\mu = \frac{E}{2(1+v)}$ and $\lambda = \frac{Ev}{(1+v)(1-2v)}$.

The sensitivity of the 2$^{nd}$ Piola-Kirchoff stress $\mathbf{S}$ to $q$-th material parameter $\mathbf{S},_{P_q}$ can be written as:

$$\mathbf{S},_\mu = \frac{\partial \mathbf{S}}{\partial \mu} = \mathbf{I} - \mathbf{C}^{-1}$$
$$\mathbf{S},_\lambda = \frac{\partial \mathbf{S}}{\partial \lambda} = (\ln J)\mathbf{C}^{-1} \tag{13}$$

Moreover, the elasticity tensor $\mathbf{K}$ can be written as [21]:

$$\mathbf{K} = 2\frac{\partial \mathbf{S}}{\partial \mathbf{C}} = \lambda(\mathbf{C})^{-1} \otimes (\mathbf{C})^{-1} + 2(\mu - \lambda \ln(J))(\mathbf{C})^{-1} \odot (\mathbf{C})^{-1} \tag{14}$$

where $(\mathbf{C})^{-1} \odot (\mathbf{C})^{-1} = -\frac{\partial \mathbf{C}^{-1}}{\partial \mathbf{C}}$ and $\otimes$ is the dyadic multiplication symbol.

**The Virtual fields method**

In this part, we use the principle of virtual power applied onto the 2$^{nd}$ Piola-Kirchhoff stress tensor (material form of the principle of virtual power) to rewrite Eq. (10) in its weak form:

$$\int_\beta \delta \mathbf{S}^o : \delta \mathbf{E}^{o(n)} dV = -\int_\beta (\delta \mathbf{F}^o \mathbf{S}^o) : \delta \mathbf{E}^{o(n)} dV \tag{15}$$

where $\delta \mathbf{E}^{o(n)}$ is the virtual Green strain field produced by a kinematically admissible virtual displacement field. The index $n$ indicates that at least $N$ virtual fields are necessary to construct the whole system of $N$ equations for solving $N$ unknown material properties $P_N$.

Next, we replace $\delta \mathbf{F}^o$ with $\delta \boldsymbol{\varepsilon}^o$ as only the symmetric parts contribute to the product. Substituting Eq. (9) into Eq. (15), we have:

$$\Sigma_{q=1}^N \delta P_q^o \int_\beta \mathbf{S}^o,_{P_q} : \delta \mathbf{E}^{o(n)} dV = -\int_\beta (\mathbf{K}^o : \delta \mathbf{E}^o + \delta \boldsymbol{\varepsilon}^o \mathbf{S}^o) : \delta \mathbf{E}^{o(n)} dV \tag{16}$$

## Construction of virtual fields

The choice of virtual fields is a critical issue in the VFM procedure. Herein we provide an efficient method of constructing $N$ virtual strain fields $\delta \mathbf{E}^{o(n)}$ for solving the inverse problem.

For convenience, we introduce the linear operator $\mathbf{L}$ which transforms a second order tensor $\mathbf{U}$ into another second order tensor $\mathbf{L}:\mathbf{U}$ such as:

$$\mathbf{L}:\mathbf{U} = \mathbf{K}^o : \mathbf{U} + (\mathbf{F}^o)^{-T} \mathbf{U}(\mathbf{F}^o)^{-1} \mathbf{S}^o \tag{17}$$

Then we use the linear operator $\mathbf{L}$ to construct the virtual fields $\delta \mathbf{E}^{o(n)}$ such as:

$$\delta \mathbf{E}^{o(n)} = (\mathbf{L})^{-1} : (\mathbf{L})^{-1} : \mathbf{S}^o_{,P_q} \tag{18}$$

Yielding,

$$\Sigma_{q=1}^{N} \delta P_q^o \int_{\beta} \left((\mathbf{L})^{-1} : \mathbf{S}^o_{,P_q}\right) : \left((\mathbf{L})^{-1} : \mathbf{S}^o_{,P_n}\right) dV = -\int_{\beta} \delta \mathbf{E}^o : \left((\mathbf{L})^{-1} : \mathbf{S}^o_{,P_n}\right) dV \tag{19}$$

Let us then introduce the following second order tensor

$$\mathbf{E}^o_{,P_n} = (\mathbf{L})^{-1} : \mathbf{S}^o_{,P_n} \tag{20}$$

Assuming that the set of $\mathbf{E}^o_{,P_n}$ tensors constitutes a basis of the space of Green-Lagrange strain fields, the whole VFM system of Eq. (16) could be written in a very concise and elegant way by substituting Eq. (20) into Eq. (19).

In the resulting system written in Eq. (21), each component $n$ of the unknown parameters variation $\{\delta P^o\}$ can be interpreted as the projection of $\delta \mathbf{E}^o$ onto the base tensor $\mathbf{E}^o_{,P_n}$, making them unique. This may be rewritten as

$$\begin{bmatrix} \int_{\beta} \mathbf{E}^o_{,P_1} : \mathbf{E}^o_{,P_1} dV & \cdots & \int_{\beta} \mathbf{E}^o_{,P_N} : \mathbf{E}^o_{,P_1} dV \\ \vdots & \ddots & \vdots \\ \int_{\beta} \mathbf{E}^o_{,P_1} : \mathbf{E}^o_{,P_N} dV & \cdots & \int_{\beta} \mathbf{E}^o_{,P_N} : \mathbf{E}^o_{,P_N} dV \end{bmatrix} \begin{pmatrix} \delta P_1^o \\ \vdots \\ \delta P_N^o \end{pmatrix} = -\begin{pmatrix} \int_{\beta} \delta \mathbf{E}^o : \mathbf{E}^o_{,P_1} dV \\ \vdots \\ \int_{\beta} \delta \mathbf{E}^o : \mathbf{E}^o_{,P_N} dV \end{pmatrix} \tag{21}$$

This equation can be summarized as $[A^o]\{\delta P^o\} = \{b^o\}$ and the material properties are updated such as $P = P^o + [A^o]^{-1}\{b^o\}$.

For a nonhomogeneous case, the domain of interest $\beta$ in Eq. (21) could be defined as each homogeneous region or even as the nodal value of each finite element node. Thus, $\{\delta P^o\}$ could be the variational parameter of a specific region or finite element node. We denote RE-VFM and NO-VFM for the regional inverse scheme and for the nodal inverse scheme, respectively. Thus, we solve $M$ times the $[A^o]\{\delta P^o\} = \{b^o\}$ equation systems ($M$ could be the total number of regions for RE-VFM and the number of finite element nodes for NO-VFM) and finally recover the entire nonhomogeneous distribution.

**Nonlinear FEM algorithm**

Our method requires achieving a number of forward finite element calculations. We performed them in FEniCS. We adopted the Newton Raphson method for solving the nonlinear variational problem [21] [22] and MUMPS (multifrontal massively parallel sparse direct solver) package [23] to solve the linear algebraic equations.

## FEniCS implementation

In this section, we elaborate step by step the novel VFM algorithms implemented in the FEniCS platform using Python scripts, and describe the corresponding `main.py` given in **Appendix** for a bilayer cubic structure.

In the proposed VFM-based inverse scheme, we construct a series of intermediate configurations for approaching the current configuration by updating the initial parameter vector $P^o$. The flowchart of the inverse algorithm is shown in **Figure 2**.

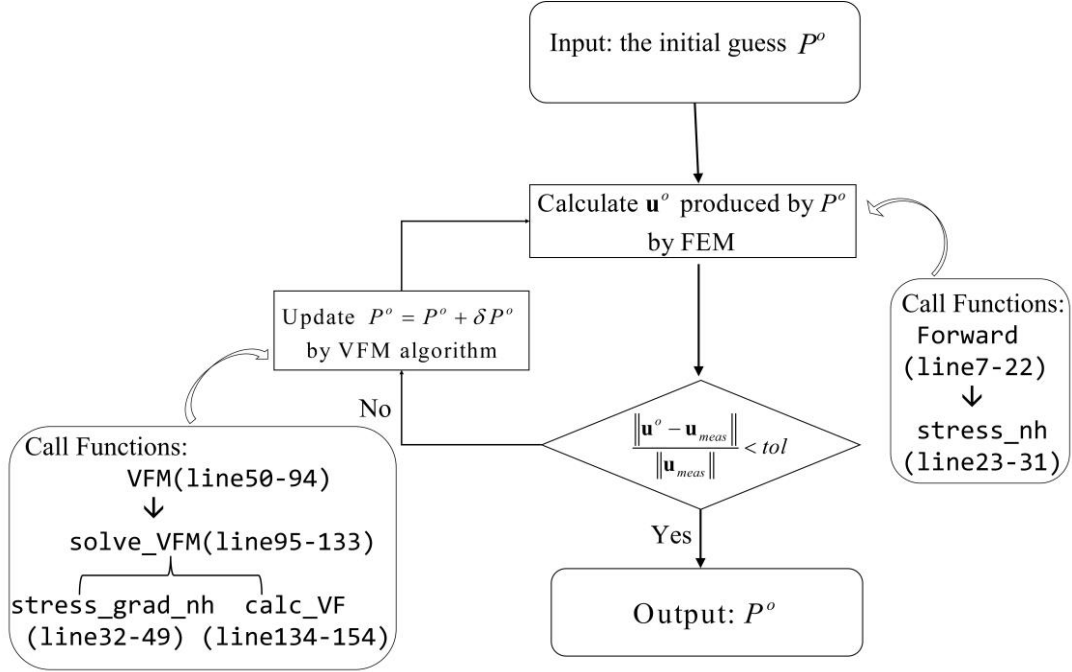

**Figure 2** Flowchart of the proposed VFM-based inversion algorithm

**FEM forward problem (lines 7-22)**

In this section, we describe the resolution of the forward problem with function `forward`. The input is the Neo-Hookean constitutive parameters $E$ and $v$, tractions and boundary conditions. In the absence of body force, the objective function of this nonlinear problem $\pi$ is written as Eq. (22). We adopt a Newton's method with a mumps solver and return the displacement field $\mathbf{u}$ in this case.

$$\pi = \int_\beta \mathbf{FS}:\nabla v^* dV - \int_{\partial \beta} \mathbf{t} \cdot v^* dS \qquad (22)$$

where $v^*$ is the test function.

**Constitutive model (lines 23-49)**

In this section, we establish the constitutive model in the `stress_nh` and `stress_grad_nh` functions. The inputs of both functions are the displacement field $\mathbf{u}$ and constitutive parameters $E$, $v$. The `stress_nh` function returns the 1st Piola-Kirchoff stress tensor in Eq. (12) for solving the forward problem. The `stress_grad_nh` function returns the sensitivity of the 2$^{nd}$ Piola-Kirchoff stress in Eq. (13) and the 4-order tensor $\mathbf{L}$ in Eq. (17) for formulating the VFM solving system in Eq. (21). Performing the tensor symbolic calculation is rather easy thanks to the FEniCS package UFL.

**VFM updating algorithm (lines 50-154)**

In this section, we introduce the VFM updating algorithm consisting of functions `VFM`, `calc_VF`, and `solve_VFM`.

Function `VFM` is the main loop function with the initial guess of material parameters as input. We set the initial guess of material parameters to the model in lines 54-59. Then, we call the `forward` function to calculate $\mathbf{u}^o$ in the intermediate configuration.

The convergence criterion of this paper is: the relative error between $\mathbf{u}^o$ and $\mathbf{u}_{meas}$ is less than $10^{-6}$ ). The error between the computed displacement field $\mathbf{u}^o$ and the measured displacement field $\mathbf{u}_{meas}$ in line 61 is defined as in Eq. (23). In lines 69-71 we calculate the deformation gradient tensor $\mathbf{F}^o$, right Cauchy-Green tensor $\mathbf{C}^o$ and the Green strain tensor $\mathbf{E}^o$ in the intermediate configuration.

$$error = \frac{\int_\beta (\mathbf{u}^o - \mathbf{u}_{meas})^2 dV}{\int_\beta (\mathbf{u}_{meas})^2 dV} \tag{23}$$

Next, we call the `stress_grad_nh` function to calculate $\mathbf{L}$, $\mathbf{S}_{,E}$ and $\mathbf{S}_{,v}$. It is not easy to obtain the inverse of a 4-order tensor in Eq. (20). To address this issue, we `project` all these tensor field to each node so that we can transform the data type from `Tensor` to `NumPy.array` in lines 74-75. In RE-VFM, we call `solve_VFM` to solve the assembled Eq. (21) and update $P^o$ in lines 77-94.

Function `clac_VF in` lines 134-154 is used to calculate the virtual fields in Eq. (20). We input the `NumPy.array` type $\mathbf{S}_{,E}$, $\mathbf{S}_{,v}$, $\mathbf{E}^o$, $\mathbf{E}_{meas}$ and $\mathbf{L}$ written in the nodal-value form and return $N$ (two in this case) virtual strain fields $\mathbf{E}^o_{,E}$, $\mathbf{E}^o_{,v}$ and the Green-Lagrange strain fields $\mathbf{E}^o$, $\mathbf{E}_{meas}$. Due to the symmetry, we rewrite the 2-order tensor from `shape([3,3])` to `(6,1)` and 4-order tensor from `shape([3,3,3,3])` to `(6,6)`.

Function `solve_VFM` in lines 95-133 is used to solve the VFM system in Eq. (21) by RE-VFM. In lines 97-104, we group the nodes into `top_dof_list` and `bottom_dof_list`, where the vertices index `0-243` belongs to the top layer and the rest belong to the upper layer. We should index the sequence by `dof2vtx`, translating the sequence arrangement from `dof` to vertices order. In lines 105-133, we assemble and solve two VFM updating systems for each layer of the bilayer structure for RE-VFM.

For NO-VFM, we can modify the `solve_VFM` function in lines 95-133 and the corresponding updating code in lines 77-93. The modified `solve_VFM` function for NO-VFM can be rewritten as:

```
1.  def solve_VFM(dS_dE_array,dS_dnu_array,E_array,E_meas_array,L_array):
2.      Beta = np.zeros([num_vertices,2])
3.      for ii in range(num_vertices):
4.          B = np.zeros([2,1])
5.          A = np.zeros([2,2])
6.          Vir_E1,Vir_E2,E_o,E_meas = calc_VF(dS_dE_array,dS_dnu_array,E_array,E_meas_array,L_array,ii)
7.          A[0,0] = np.dot(Vir_E1.T,Vir_E1)[0][0]
8.          A[0,1] = np.dot(Vir_E2.T,Vir_E1)[0][0]
9.          A[1,0] =np.dot(Vir_E1.T,Vir_E2)[0][0]
10.         A[1,1] = np.dot(Vir_E2.T,Vir_E2)[0][0]
11.         B[0] = -np.dot((E_meas-E_o).T,Vir_E1)[0][0]
12.         B[1] = -np.dot((E_meas-E_o).T,Vir_E2)[0][0]
13.         cond = np.linalg.cond(A)
14.         if cond>=1e6:
15.             temp_beta = np.linalg.lstsq(A,B)[0]
16.         else:
17.             temp_beta = np.linalg.solve(A,B)
18.         Beta[ii,0] = temp_beta[0]
19.         Beta[ii,1] = temp_beta[1]
20.     return Beta
```

**Main part (lines 155-213)**

In this section, we elaborate the entire procedure of the inverse algorithm. In lines 156-

160, we import the geometry mesh file. In lines 162-172, we define functions and function spaces in FEniCS. In lines 173-186, we apply the displacement and traction boundary conditions.

Next, we use a finite element simulation as experimental measurement for the sake of verification. In lines 187-195, we define the target distribution of Young's modulus and Poisson's ratio for the finite element model. From lines 196-201, we calculate $\mathbf{u}_{meas}$, $\mathbf{F}_{meas}$, $\mathbf{C}_{meas}$ and $\mathbf{E}_{meas}$ successively in the current configuration.

In lines 202-209, we make preparations for the VFM iteration and set the initial guess of parameters in the intermediate configuration. In lines 211-213, we call the VFM algorithm to finish the identification and post-process the reconstructed results by `post_plot`. The identification results of the cases are discussed in the **Results** section in details.

## Results

**Bilayer Structure Problem**

We first considered a bilayer structure as shown in **Figure 3**, the model is divided into $9 \times 9 \times 5$ nodes and 1280 tetrahedrons. We fixed the bottom surface and applied the constant traction on the top surface of the model. The whole model was equally divided into two layers. The target Young's moduli of the top layer and the bottom layer were 10MPa and 20MPa, respectively. The Poisson's ratio was 0.3 for both layers. Firstly, we applied the RE-VFM to solve this identification problem. In this case, the initial guess of Young's moduli and Poisson's ratios of both the top layer and bottom layer are 15 MPa and 0.2, respectively. The estimated material properties and the error of the displacement fields with respect to the iteration number are plotted in **Figure 4** and **Figure 5**, respectively. The relative error between the estimated and target material properties is reported in **Table 1**. We observed that both values of Young's modulus and Poisson's ratio were identified accurately. Moreover, the displacement error is less than the tolerance after 16 iterations.

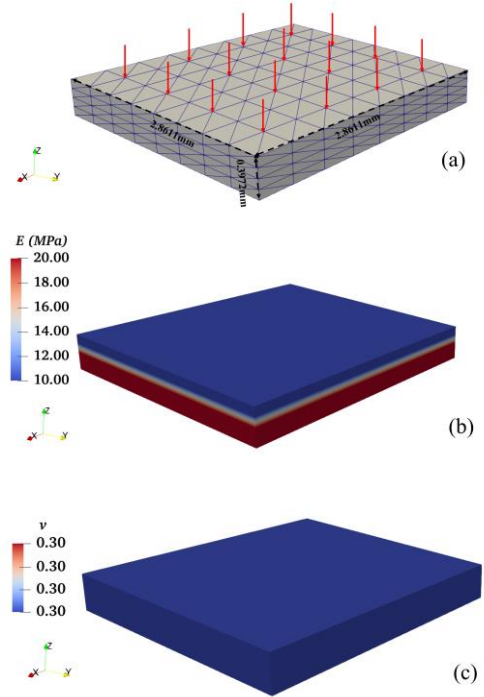

**Figure 3** The finite element bilayer structure model and target material parameter distributions. (a) the finite element model and the boundary conditions; (b) target Young's modulus distribution; (c) target distribution of the Poisson's ratio.

**Table 1** Relative error of the estimated parameters of the bilayer model

|  | Target | Initial guess | Estimation | Relative error |
| --- | --- | --- | --- | --- |
| $E_{top}$ | 10.00 (MPa) | 15.00 (MPa) | 9.9675 (MPa) | 0.32% |
| $E_{bottom}$ | 20.00 (MPa) | 15.00 (MPa) | 20.0440 (MPa) | 0.22% |
| $v_{top}$ | 0.30 | 0.20 | 0.3015 | 0.49% |
| $v_{bottom}$ | 0.30 | 0.20 | 0.2992 | 0.27% |

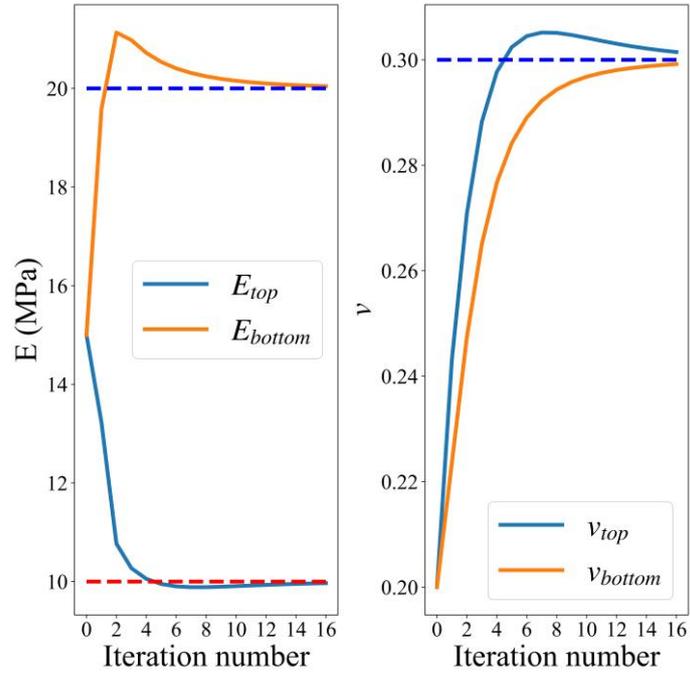

**Figure 4** Estimated values of material properties versus the iteration number.

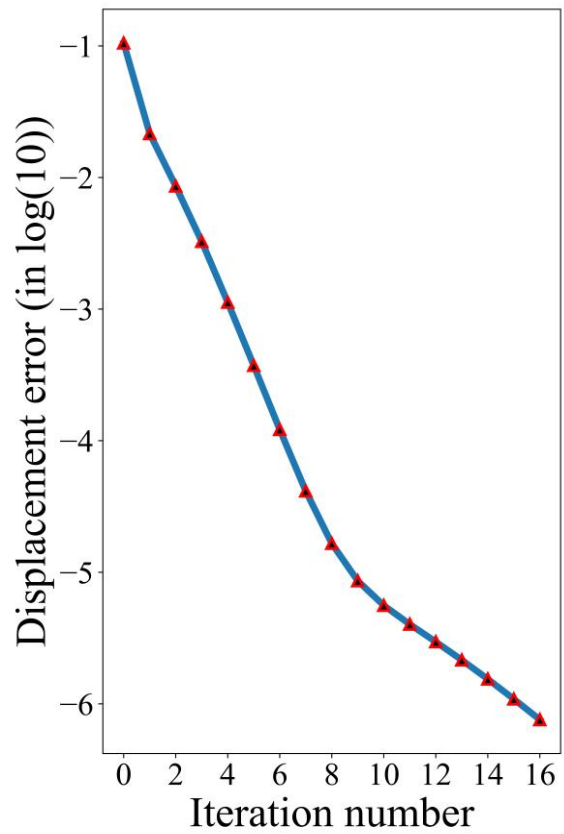

**Figure 5.** Error between $\mathbf{u}_{meas}$ and $\mathbf{u}^o$ of the bilayer case by RE-VFM. (The log10 of the error is adopted)

Subsequentially, we tested the NO-VFM method with the bilayer structure case. The initial guess of Young's moduli and Poisson's ratios for all nodes are set to 15 MPa and 0.2, respectively. In **Figure 6** and **Figure 7**, we present the reconstructed results and the displacement error of the bilayer case by the NO-VFM method. Since this is a large-scale inverse problem, more minimization iteration numbers are required to satisfy the convergence criteria. Additionally, both the Young's modulus and Poisson's ratio distributions were well recovered, as shown in **Figure 6**. The average relative error of the Young's moduli and Poisson's ratios is 11.62% and 4.51%, respectively.

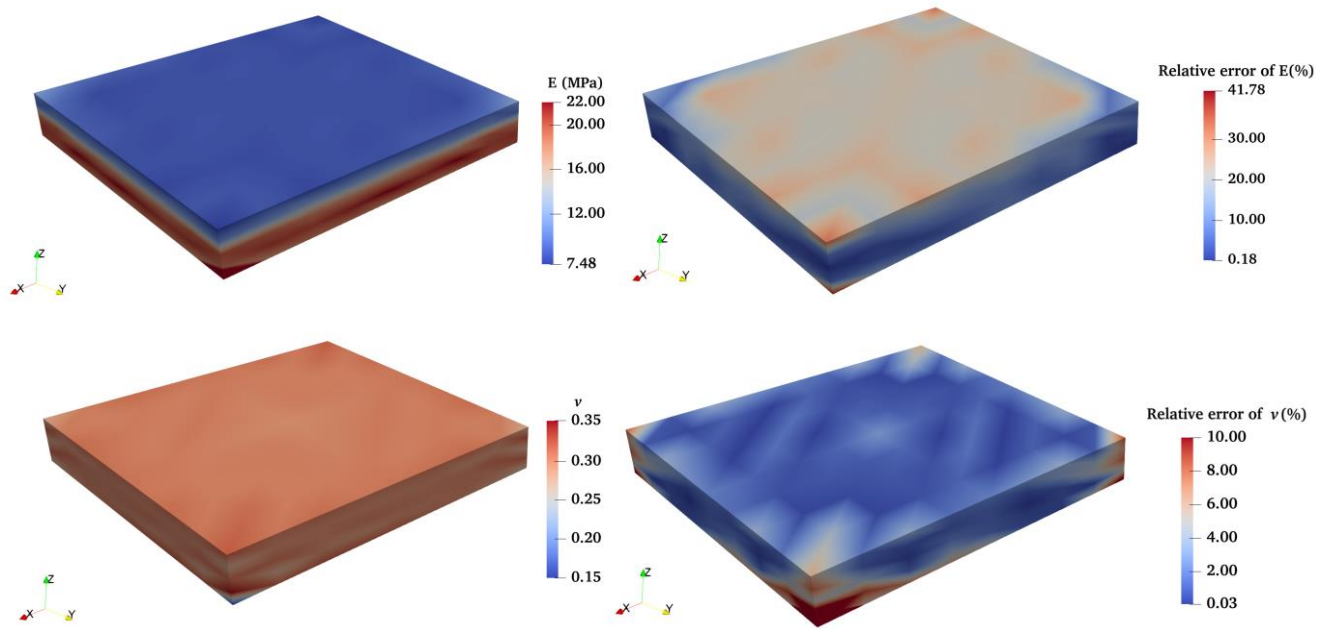

**Figure 6** Reconstructed results of the bilayer case by NO-VFM. 1st row: the reconstruction result and relative error distribution of Young's modulus; 2nd row: the reconstruction result and relative error distribution of Poisson's ratio.

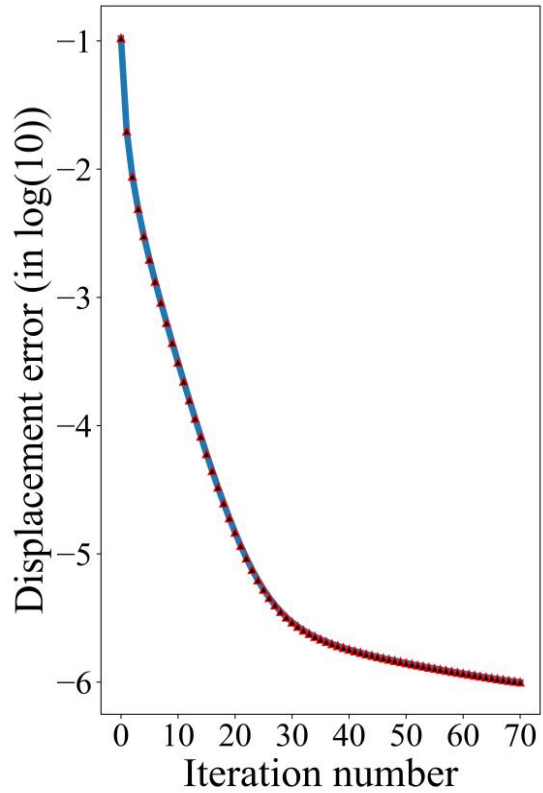

**Figure 7** Error between $\mathbf{u}_{meas}$ and $\mathbf{u}^o$ of the bilayer case by NO-VFM. (The log10 of the error is adopted)

Next, we test the situation when the initial guesses are far from the target values for both RE-VFM and NO-VFM using the bilayer model. For RE-VFM, the initial guess of Young's moduli and Poisson's ratios of both the top layer and bottom layer are set to 1 MPa and 0.2, respectively. The relative error between the estimated and target material properties is reported in **Table 2**. The estimated material properties and the error of the displacement fields with respect to the iteration number are plotted in

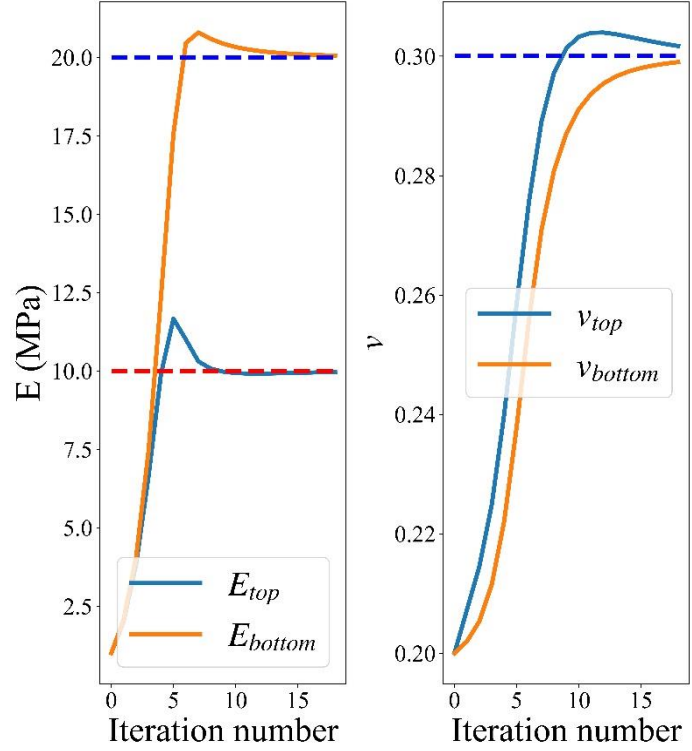

**Figure 8** and **Figure 9**, respectively. We observe that the accurate values are obtained after 19 iterations, even the initial guess is far away from the target parameter values.

**Table 2** Relative error of the estimated parameters of the bilayer model

|  | Target | Initial guess | Estimation | Relative error |
|---|---|---|---|---|
| $E_{top}$ | 10.00 (MPa) | 1.00 (MPa) | 9.9627 (MPa) | 0.37% |
| $E_{bottom}$ | 20.00 (MPa) | 1.00 (MPa) | 20.0522 (MPa) | 0.26% |
| $v_{top}$ | 0.30 | 0.20 | 0.3017 | 0.56% |
| $v_{bottom}$ | 0.30 | 0.20 | 0.2990 | 0.33% |

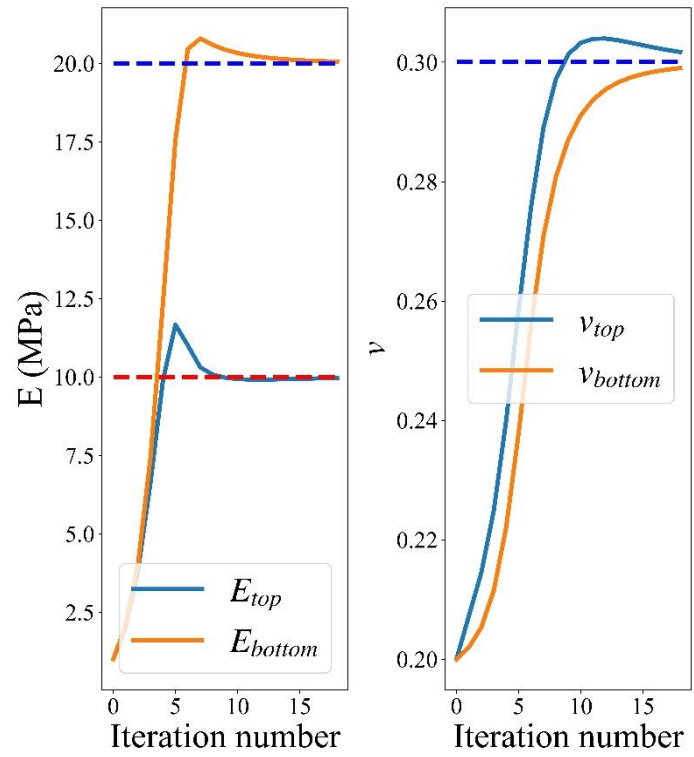

**Figure 8** Estimated values of material properties versus the iteration number.

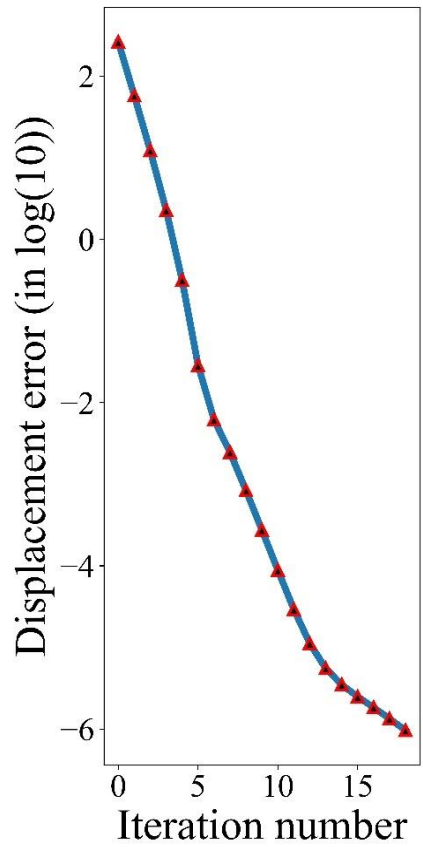

**Figure 9** Error between $\mathbf{u}_{meas}$ and $\mathbf{u}^o$ of the bilayer case by RE-VFM. (The log10 of the error is adopted)

In NO-VFM, the initial guess of Young's moduli and Poisson's ratios for all nodes are set to 1 MPa and 0.2, respectively. In **Figure 10** and **Figure 11**, we present the reconstructed results and the displacement error. As shown in **Figure 10**, both the Young's modulus and Poisson's ratio distributions were well recovered even the initial guesses are far away from the target values. After 71 iterations, the average relative error of the Young's moduli and Poisson's ratios are 9.89% and 5.97%, respectively.

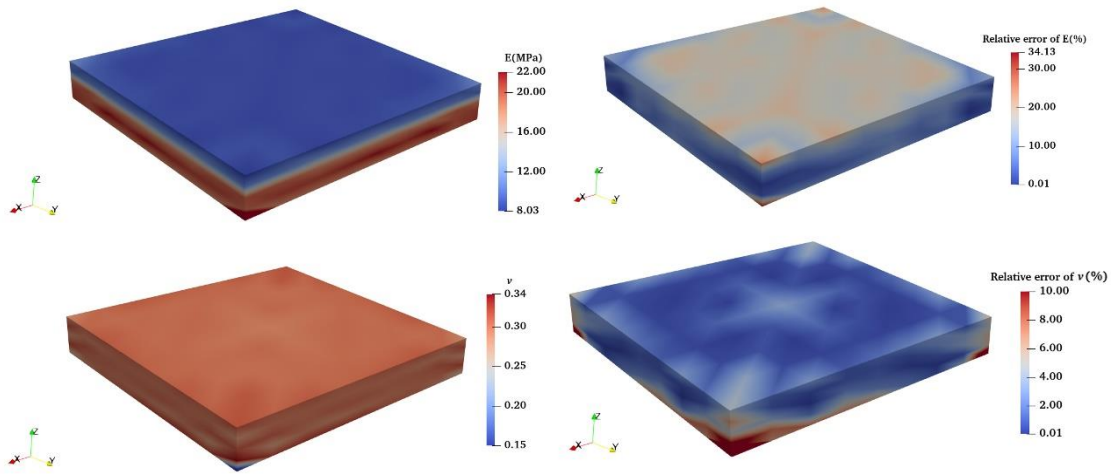

**Figure 10** Reconstructed results of the bilayer case by NO-VFM. 1st row: the reconstruction result and relative error distribution of Young's modulus; 2nd row: the reconstruction result and relative error distribution of Poisson's ratio.

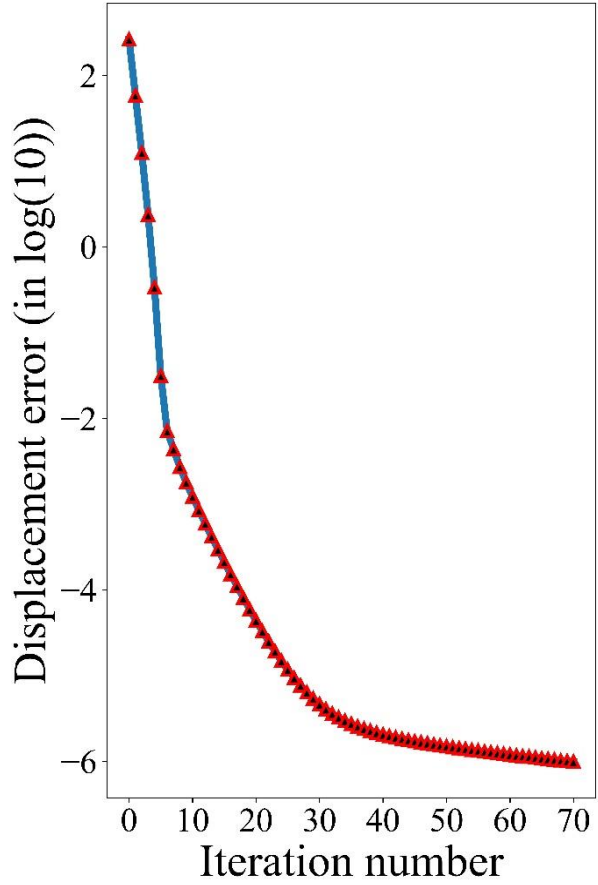

**Figure 11** Error between $\mathbf{u}_{meas}$ and $\mathbf{u}^o$ of the bilayer case by NO-VFM. (The log10 of the error is adopted)

**Lamina Cribosa (LC) Model**

Next, we tested the NO-VFM algorithm for a three-layered lamina cribosa (LC) model, with 6581 nodes and 33451 tetrahedral elements. Lamina cribosa is a connective tissue in the optic nerve head (ONH), whose mechanical properties could provide important information of studying the cause of glaucoma. This geometric model was previously used for other studies [14][24]. As boundary conditions, we fixed the posterior and lateral surfaces and applied pressure on the upper surface. We assumed that the LC was nearly incompressible and the Poisson's ratio was set as 0.46. The target Young's modulus distribution is shown in **Figure 12**. The initial guess of Young's modulus is 0.4 MPa for all nodes. The mean relative error between the target and recovered Young's moduli is 0.11%. In **Figure 13**, we observed that the three-layer structure was well reconstructed by the NO-VFM. In **Figure 14**, the error of the displacement fields reduced to 10$^{-6}$ after 5 iterations.

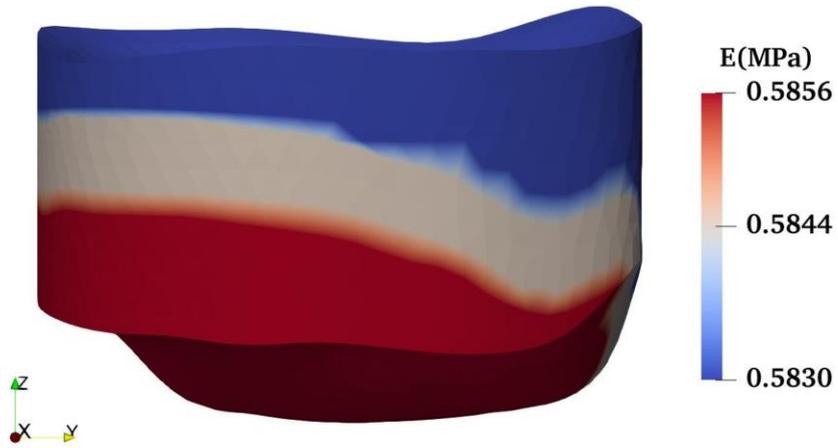

**Figure 12** Target Young's modulus distribution of the LC case.

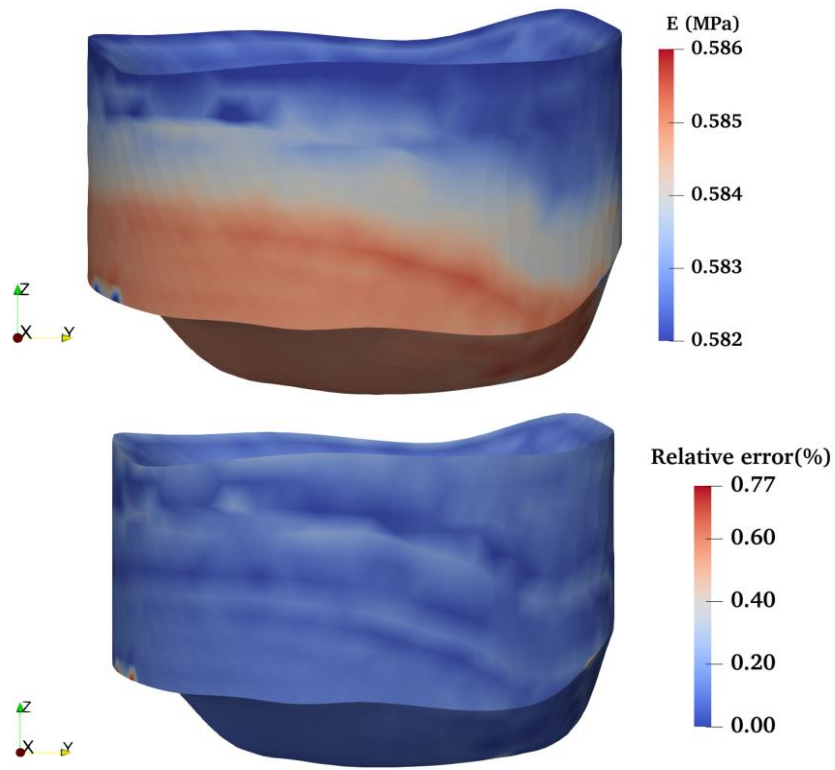

**Figure 13** Reconstructed results of the LC case by NO-VFM

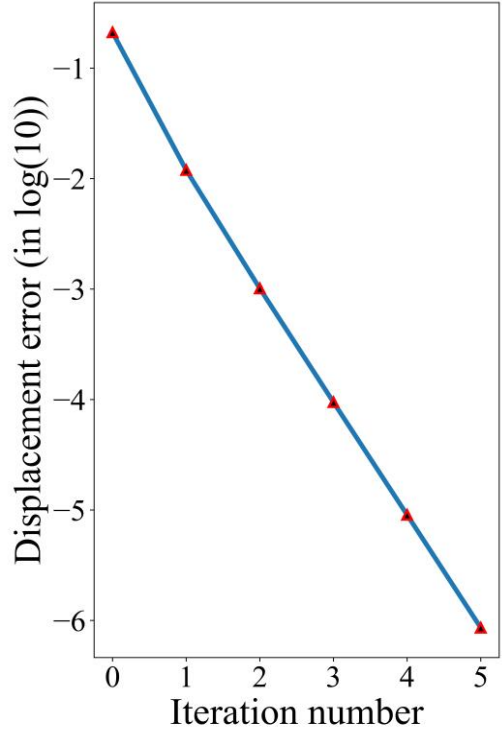

**Figure 14** Error between $\mathbf{u}_{meas}$ and $\mathbf{u}^o$ of the LC case by NO-VFM. (The log10 of the error is adopted)

**Inclusion Problem**

Lastly, we tested the NO-VFM for an 1mm×1mm×1mm cubic model with a spherical inclusion (radius is 0.3mm) in the center. The geometric model is shown in **Figure 15**, which is discretized with 8092 tetrahedral elements. We applied the pressure on the top surface and fixed the bottom surface to avoid rigid motion. The target distributions of Young's modulus and Poisson's ratio are shown in **Figure 15**. The target Young's modulus values are 5 MPa for the inclusion and 1 MPa for the background, respectively. The target Poisson's ratio is 0.45 for the inclusion and 0.35 for the background, respectively. The initial guesses of Young's modulus and Poisson's ration are set as 1 MPa and 0.4 for all nodes. The mean relative error between the target and recovered Young's moduli is 8.60%. Moreover, the mean relative error between the target and recovered Poisson's ratios is 3.41%. We observed that the inclusion was well reconstructed by the NO-VFM method (see **Figure 16**) and the relative error reduced to
$10^{-6}$ after 45 iterations (see **Figure 17**).

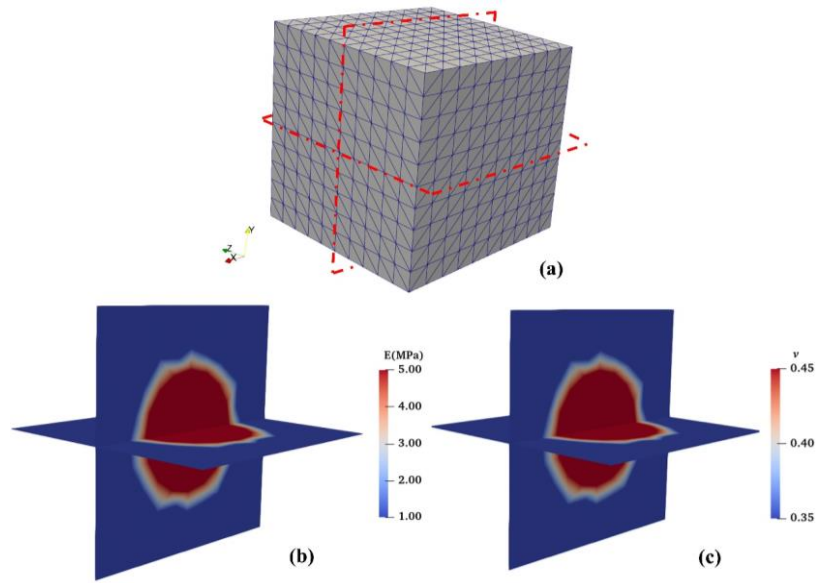

**Figure 15** The finite element model of the inclusion problem and target material parameter distributions. (a) the finite element model; (b) target distribution of Young's modulus; (c) target distribution of the Poisson's ratio.

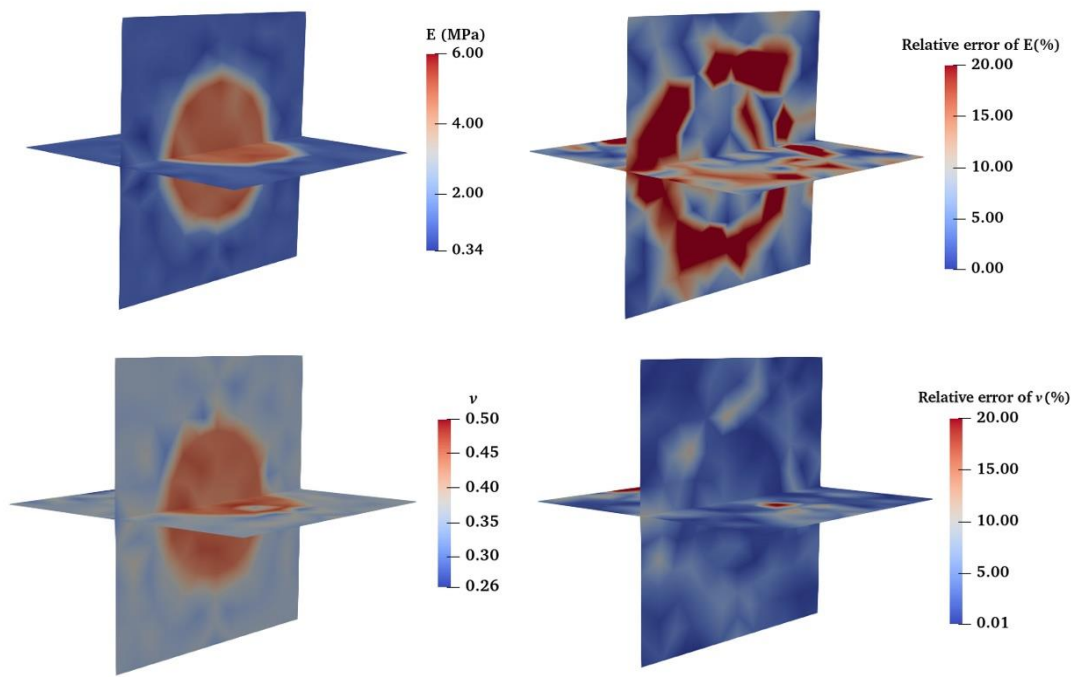

**Figure 16** Reconstructed results of the inclusion problem by NO-VFM. 1$^{st}$ row: the reconstruction result and relative error distribution of Young's modulus; 2$^{nd}$ row: the reconstruction result and relative error distribution of Poisson's ratio.

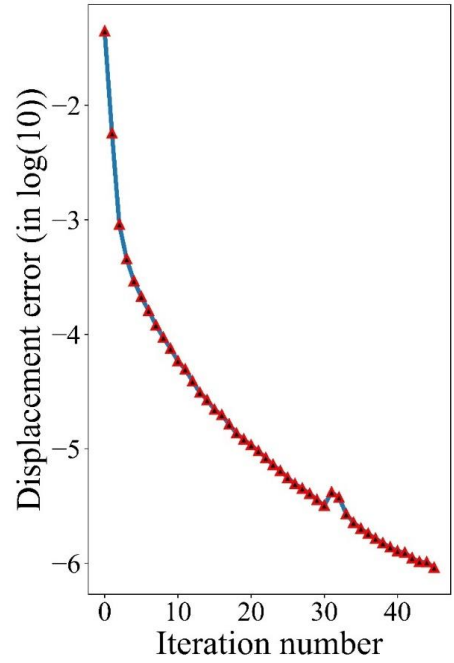

**Figure 17** Error between $\mathbf{u}_{meas}$ and $\mathbf{u}^o$ of the inclusion problem by NO-VFM. (The log10 of the error is adopted)

We ran all these examples with a personal Laptop (Intel 11800H CPU, with 8 cores 16 threads, 4.2 GHz, 32G RAM). The FEniCS environment was configured in Python 3.8 on Ubuntu18.04 (WSL, Windows Subsystem for Linux). The average computational time of the forward, transformation and VFM parts are reported in **Table 3**, respectively. **Table 3** demonstrates that most of the computation time is used to transform the `Tensor` to `Numpy`.

**Table 3** Average computational times of an iteration step

| Model | Algorithm | Forward | Transformation | VFM | Total |
| --- | --- | --- | --- | --- | --- |
| Bilayer | RE-VFM | 0.5s | 20.2s | 0.4s | 21.1s |
| | NO-VFM | 0.5s | 19.5s | 0.2s | 20.2s |
| LC | NO-VFM | 144.2s | 485.2s | 27.8s | 657.2s |
| Inclusion | NO-VFM | 9.1s | 185.4s | 0.2s | 194.7s |

## Discussion

In this paper, we have implemented two VFM methods in the open-source finite element package FEniCS. Choosing FEniCS was motivated by the relative simplicity of implementing a finite element code with the intrinsic high-level language in FEniCS. This makes the proposed method highly flexible and convenient to introduce user-defined material constitutive models.

In this study, the isotropic Neo-Hookean hyperelastic constitutive model was chosen to test the proposed VFM-based inverse methods. Numerical examples have shown the feasibility of the proposed methods, even when the initial guesses are far from the target values. However, we should also test the feasibility of the proposed methods by other types of constitutive models considering exponentially nonlinear, anisotropic and viscoelastic behaviors.

A recurrent difficulty in inverse method is related to the uncertainty about the boundary conditions. The inaccurate boundary might affect the accuracy of the identification results. In this paper, the traction **t** is applied in the reference configuration. Eq 5 is the quasi-static equilibrium equation of forces responsible for deformation **F** and Eq 6 is the quasi-static equilibrium equation of forces responsible for deformation $\mathbf{F}^\circ$, both written in the reference configuration. As they are written in the reference configuration, both equations have the same **t** traction vector. This means that there are no assumptions on the boundary conditions in our method.

A bottleneck of VFM methods compared to the optimization-based methods is that artificial virtual fields should be introduced. With the increasing number of unknown material parameters, the total number of virtual fields should be increased accordingly to ensure the uniqueness of the inverse solution. To address this issue, RE-VFM method was firstly proposed in [14] and can be used to solve for the regional material properties without introducing artificial virtual fields. However, the virtual fields in [14] are obtained by solving another constructed forward problem, which requires more time. In this paper, a novel way to construct virtual fields is proposed. Based on the proposed method, we further develop the NO-VFM method, which is capable of identifying the spatial variation of the nonhomogeneous elastic property distribution of soft solids. Compared to [15], the NO-VFM method can be used to identify the material properties for a hyperelastic solid with a complex geometry. The feasibility of the proposed method has been successfully tested by several numerical examples were shown in the **Results** Section. The simulated displacement datasets utilized to test the feasibility of the proposed VFM methods illustrate different possible applications. In the future, the performance of the proposed methods should be tested with actual datasets obtained from medical images.

## Conclusions

In this study, we have proposed two efficient VFM methods for identifying the distribution of hyperelastic material parameters in soft tissues. The details of the implementation of the proposed methods in the FEniCS platform were presented. From the observation of the numerical examples, we have found that both the RE-VFM and NO-VFM methods are capable of reconstructing the hyperelastic nonhomogeneous material distribution accurately. In particular, the NO-VFM shows the potential of reconstructing the material distribution without the partition information. This study has demonstrated that the proposed methods have great potential in identifying regional variations of material properties in soft tissues due to diseases and aging.

## Acknowledgements

The authors acknowledge the support from the National Natural Science Foundation of China (11732004, 12002075), 111 Project (B14013), National Key Research and Development Project (2020YFB1709401), Natural Science Foundation of Liaoning Province in China(2021-MS-128).

We are very garteful to Prof. Thao D. Nguyen and Dr. Brandon Zimmerman from Johns Hopkins University for sharing the lamina cribosa model and for the useful discussions

# Appendix

We present the `main.py` and `util.py` files of the double-layered structural model below. `main.py` is utilized to implement the VFM-based inverse algorithm. `util.py` is used to store the auxiliary code pieces such as `project` and `post_plot`.

`main.py`

```python
#%%  VFM-based inverse algorithm by FeniCS2019
from fenics import *
import ufl as uf
import os
import numpy as np
from utils import proj_tensor2,proj_tensor4,post_plot
def forward(E,nu,Trac,bcs):
    P= stress_nh(u,E,nu)
    FV = inner(P, grad(v))*dx
    # Traction at boundary
    T=Trac
    FT=dot(T,v)*ds1
    # Whole system and its Jacobian
    FF = FV-FT
    JJ = derivative(FF, u)
    # Initialize solver
    problem = NonlinearVariationalProblem(FF, u, bcs=bcs, J=JJ)
    solver = NonlinearVariationalSolver(problem)
    solver.parameters['newton_solver']['relative_tolerance'] = 1e-10
    solver.parameters['newton_solver']['linear_solver'] = 'mumps'
    solver.solve()
    return u
def stress_nh(u,E,nu):
    """Returns 1st Piola-Kirchhoff stress and (local) mass balance for given u, p."""
    mu = E/(2.0*(1.0 + nu)); lam= E*nu/((1.0 + nu)*(1.0 - 2.0*nu))
    F = I + grad(u)
    J = det(F)
    C = F.T * F
    S = mu*(I-inv(C))+lam*ln(J)*inv(C) # 2nd Piola-Kirchoff stress
    P = F*S # 1st Piola-Kirchhoff stress
    return P
def stress_grad_nh(u,E,nu):
    i,j,k,l,m=uf.indices(5)
    mu = E/(2.0*(1.0 + nu))
    lam= E*nu/((1.0 + nu)*(1.0 - 2.0*nu))
    dmu_dE = 1.0/(2.0*(1.0 + nu))
    dlam_dE= 1.0*nu/((1.0 + nu)*(1.0 - 2.0*nu))
    dmu_dnu = -1.0*E/(2.0*(1.0 + nu)*(1.0 + nu))
    dlam_dnu= E*(2.0*nu*nu+1.0)/((1.0 + nu)*(1.0 - 2.0*nu)) /((1.0 + nu)*(1.0 - 2.0*nu))
    F = I + grad(u)
    J = det(F)
    C=(F.T)*F
    C_inv=inv(C)
    S = mu*(I-inv(C))+lam*ln(J)*inv(C) # 2nd Piola-Kirchhoff stress
    dS_dE=dmu_dE*(I-inv(C))+dlam_dE*ln(J)*inv(C) # 2nd Piola-Kirchhoff stress
    dS_dnu=dmu_dnu*(I-inv(C))+dlam_dnu*ln(J)*inv(C) # 2nd Piola-Kirchhoff stress
    K_tan=as_tensor(lam*C_inv[i,j]*C_inv[k,l]+(mu-lam*ln(J))*(C_inv[i,k]*C_inv[j,l]+C_inv[i,l]*C_inv[j,k]),(i,j,k,l))
    L = as_tensor(K_tan[i,j,k,l]+inv(F)[k,i]*inv(F)[l,m]*S[m,j],(i,j,k,l))
    return dS_dE,dS_dnu,L
def VFM(E_top_gue,E_bottom_gue,nu_top_gue,nu_bottom_gue):
    u_o = Function(V)
    iter = 0
    while True:
        E_top_list.append(E_top_gue);E_bottom_list.append(E_bottom_gue)
        nu_top_list.append(nu_top_gue);nu_bottom_list.append(nu_bottom_gue)
        E_o = interpolate(Expression('x[2]>-0.19635? E_top:E_bottom',degree=1,E_top=E_top_gue,E_bottom=E_bottom_gue), M)
        nu_o = interpolate(Expression('x[2]>-0.19635? nu_top:nu_bottom',degree=1,nu_top=nu_top_gue,nu_bottom=nu_bottom_gue), M)
        File("output/{}/{}/Parameters/Eo_iter{}.pvd".format(case_name,initial_name,iter)) << E_o
        File("output/{}/{}/Parameters/nuo_iter{}.pvd".format(case_name,initial_name,iter)) << nu_o
        u_o.assign(forward(E_o,nu_o,traction1,bcs))
```

```python
61.            error = assemble((inner(u_o-u_meas,u_o-u_meas)*dx))/assemble((inner(u_meas,u_meas)*dx))
62.            error_list.append(error)
63.            if error < tol:
64.                break
65.            if iter >MAX_ITER:
66.                break
67.            iter += 1
68.            # kinematics in the intermediate configuration
69.            F=I+grad(u_o)
70.            C=(F.T)*F
71.            Strain_E = (C-I)/2
72.            # formulating Eq.(21)
73.            dS_dE,dS_dnu,L = stress_grad_nh(u_o,E_o,nu_o)
74.            dS_dE_array,dS_dnu_array,Strain_E_array = proj_tensor2(dS_dE,dS_dnu,Strain_E,TT)
75.            L_array = proj_tensor4(L,TT_4)
76.            # solve VFM equation systems
77.            dE_top,dE_bottom,dnu_top,dnu_bottom  = solve_VFM(dS_dE_array,dS_dnu_array,Strain_E_array,E_meas_array,L_array)
78.            E_top_gue += dE_top
79.            E_bottom_gue += dE_bottom
80.            nu_top_gue += dnu_top
81.            nu_bottom_gue += dnu_bottom
82.            if E_top_gue <= 0:
83.                E_top_gue = 0.1
84.            if nu_top_gue<=0:
85.                nu_top_gue = 0.2
86.            if nu_top_gue >=0.5:
87.                nu_top_gue = 0.48
88.            if E_bottom_gue <= 0:
89.                E_bottom_gue = 0.1
90.            if nu_bottom_gue<=0:
91.                nu_bottom_gue = 0.2
92.            if nu_bottom_gue >=0.5:
93.                nu_bottom_gue = 0.48
94.            print('This is the {} iter, error is {}'.format(iter,error))
95. def solve_VFM(dS_dE_array,dS_dnu_array,E_array,E_meas_array,L_array):
96.     # turn the node list aranged by dof to node list aranged by nodes
97.     top_dof_list = []
98.     bottom_dof_list = []
99.     for i in range(num_vertices):
100.        node_index =  dof2vtx[i]
101.        if node_index < 243:
102.            top_dof_list.append(i)
103.        else:
104.            bottom_dof_list.append(i)
105.    def solve_system(part_list):
106.        """ part_list is all nodal index of a specific part """
107.        A_11_total = 0; A_12_total = 0; A_21_total = 0; A_22_total = 0
108.        B_1_total = 0; B_2_total =0
109.        for el in part_list:
110.            B = np.zeros([2,1])
111.            A = np.zeros([2,2])
112.            Vir_E1,Vir_E2,E_o,E_meas = calc_VF(dS_dE_array,dS_dnu_array,E_array,E_meas_array,L_array,el)

113.            A[0,0] = np.dot(Vir_E1.T,Vir_E1)[0][0]
114.            A[0,1] = np.dot(Vir_E2.T,Vir_E1)[0][0]
115.            A[1,0] = np.dot(Vir_E1.T,Vir_E2)[0][0]
116.            A[1,1] = np.dot(Vir_E2.T,Vir_E2)[0][0]
117.            B[0] = -np.dot((E_meas-E_o).T,Vir_E1)[0][0]
118.            B[1] = -np.dot((E_meas-E_o).T,Vir_E2)[0][0]
119.            A_11_total += A[0,0];A_12_total+=A[0,1];A_21_total+=A[1,0];A_22_total+=A[1,1]
120.            B_1_total+=B[0];B_2_total+=B[1]
121.        A_total = np.zeros([2,2]); B_total = np.zeros([2,1])
122.        A_total[0,0] = A_11_total;A_total[0,1] = A_12_total;A_total[1,0] = A_21_total;A_total[1,1] = A_22_total
123.        B_total[0] = B_1_total;B_total[1] = B_2_total
124.        cond = np.linalg.cond(A_total)
125.        if cond>=1e6:
126.            temp_beta = np.linalg.lstsq(A_total,B_total)[0]
127.        else:
128.            temp_beta = np.linalg.solve(A_total,B_total)
129.        return temp_beta[0,0],temp_beta[1,0]
130.    dE_top,dnu_top = solve_system(top_dof_list)
131.    dE_bottom,dnu_bottom = solve_system(bottom_dof_list)
132.
133.    return dE_top,dE_bottom,dnu_top,dnu_bottom
134. def calc_VF(dS_dE_array,dS_dnu_array,E_array,E_meas_array,L_array,index):
135.    dS_dE_np = ((dS_dE_array[index].reshape([3,3])))
136.    dS_dnu_np = ((dS_dnu_array[index].reshape([3,3])))
137.    E = E_array[index].reshape([3,3])
138.    E_m = E_meas_array[index].reshape([3,3])
139.    L_np = (L_array[index].reshape([3,3,3,3]))
140.    index2D1=[0,1,2,1,0,0];index2D2=[0,1,2,2,2,1]
141.    S_E=np.zeros([6,1]);S_nu=np.zeros([6,1]);E_meas=np.zeros([6,1]);E_o=np.zeros([6,1]);LL=np.zeros([6,6])
142.    for i in range(6):
143.        for j in range(0,3):
144.            LL[i,j] = L_np[index2D1[i],index2D2[i],index2D1[j],index2D2[j]]
```

```python
145.          for j in range(3,6):
146.              LL[i,j] = 2*L_np[index2D1[i],index2D2[i],index2D1[j],index2D2[j]]
147.          S_E[i] = dS_dE_np[index2D1[i],index2D2[i]]
148.          S_nu[i] = dS_dnu_np[index2D1[i],index2D2[i]]
149.          E_o[i] = E[index2D1[i],index2D2[i]]
150.          E_meas[i] = E_m[index2D1[i],index2D2[i]]
151.      inv_LL = np.linalg.inv(LL)
152.      Vir_E1 = np.dot(inv_LL,S_E)
153.      Vir_E2 = np.dot(inv_LL,S_nu)
154.      return Vir_E1,Vir_E2,E_o,E_meas
155. #%% Model setup
156. case_name = 'Case_2layer'
157. mesh_path = 'model/2layer'
158. meshfile = '2layer.xml'
159. # Geometry mesh
160. mesh = Mesh(os.path.join(os.curdir,mesh_path,meshfile))
161. num_vertices = mesh.num_vertices()
162. #%% FEniCS Functionspaces
163. V = VectorFunctionSpace(mesh, 'P', 1)
164. u = Function(V)
165. v = TestFunction(V)
166. M = FunctionSpace(mesh, "CG", 1)
167. TT = TensorFunctionSpace(mesh,'P',1)
168. shape = 4*(mesh.geometry().dim(),)
169. TT_4 = TensorFunctionSpace(mesh,'P',1,shape = shape)
170. I = Identity(3)
171. dof = vertex_to_dof_map(M)
172. dof2vtx = vertex_to_dof_map(M).argsort()
173. # Boundary definition
174. boundary_parts = MeshFunction('size_t', mesh, mesh.topology().dim()-1)
175. bottom  = AutoSubDomain(lambda x: near(x[2], -0.3972))
176. top = AutoSubDomain(lambda x: near(x[2], 0))
177. bottom.mark(boundary_parts, 1)
178. top.mark(boundary_parts, 2)
179. dx = Measure("dx",mesh)
180. ds1 = Measure("ds", mesh,subdomain_data=boundary_parts, subdomain_id=2)
181. bc0 = DirichletBC(V.sub(0), Constant(0), boundary_parts, 1)
182. bc1 = DirichletBC(V.sub(1), Constant(0), boundary_parts, 1)
183. bc2 = DirichletBC(V.sub(2), Constant(0), boundary_parts, 1)
184. bcs = [bc0,bc1,bc2]
185. normal_vector = FacetNormal(mesh)
186. traction1 = Constant((0.0, 0.0, -0.1))
187. #%% Synthetic Umeas
188. E_target = Function(M)
189. nu_target = Function(M)
190. u_meas = Function(V)
191. # Set target parameter value by Expression
192. E_top = 10; E_bottom = 20
193. E_target = interpolate(Expression('x[2]>-0.19635? E_top:E_bottom',degree=1,E_top=E_top,E_bottom=E_bottom), M)
194. nu_top = 0.3; nu_bottom = 0.3
195. nu_target = interpolate(Expression('x[2]>-
     0.19635? nu_top:nu_bottom',degree=1,nu_top=nu_top,nu_bottom=nu_bottom), M)
196. u_meas.assign(forward(E_target,nu_target,traction1,bcs))
197. F_meas=I+grad(u_meas)
198. C_meas=(F_meas.T)*F_meas
199. E_meas=(C_meas-I)/2
200. E_meas_proj =  project(E_meas,TT)
201. E_meas_array = np.array(E_meas_proj.vector()).reshape([-1,9])
202. #%% Intermediate configurations
203. E_top_gue=15; E_bottom_gue=15
204. nu_top_gue=0.2; nu_bottom_gue=0.2
205. E_top_list = [];E_bottom_list = []
206. nu_top_list = [];nu_bottom_list = []
207. initial_name = 'E_top{:.2f}_E_bot{:.2f}_nu_top{:.2f}_nu_bot{:.2f}'.format(E_top_gue,E_bottom_gue,nu_top_gue,nu
     _bottom_gue)
208. error_list = []
209. MAX_ITER = 100; tol = 1e-6
210. ## VFM iteration loop
211. VFM(E_top_gue,E_bottom_gue,nu_top_gue,nu_bottom_gue)
212. #%% post_plot
213. post_plot(case_name,initial_name,E_top_list,E_bottom_list,nu_top_list,nu_bottom_list,error_list)
214.
```

## util.py

```python
from fenics import *
import numpy as np
import os
import matplotlib.pyplot as plt
class PinPoint(SubDomain):
    def __init__(self,p):
        self.p = p
        SubDomain.__init__(self)
    def inside(self, x, on_boundary):
        return np.linalg.norm(x-self.p) < DOLFIN_EPS
class K(UserExpression):
    def __init__(self, marker_domain, k_0, k_1,k_2, **kwargs):
        super().__init__(**kwargs)
        self.marker_domain = marker_domain
        self.k_0 = k_0
        self.k_1 = k_1
        self.k_2 = k_2

    def eval_cell(self, values, x, cell):
        if self.marker_domain[cell.index] ==0:
            values[0] = self.k_0
        elif self.marker_domain[cell.index] ==1:
            values[0] = self.k_1
        else:
            values[0] = self.k_2

    def value_shape(self):
        return ()

def proj_tensor2(dS_dE,dS_dnu,E,TT):
    ## 2 order Tensor
    dS_dE_proj = project(dS_dE,TT)
    dS_dE_array = np.array(dS_dE_proj.vector()).reshape([-1,9])
    dS_dnu_proj = project(dS_dnu,TT)
    dS_dnu_array = np.array(dS_dnu_proj.vector()).reshape([-1,9])
    E_proj =  project(E,TT)
    E_array = np.array(E_proj.vector()).reshape([-1,9])
    return dS_dE_array,dS_dnu_array,E_array
def proj_tensor4(L,TT_4):
    L_proj = project(L,TT_4,solver_type='cg')
    L_array = np.array(L_proj.vector()).reshape([-1,81])
    return L_array

def post_plot(case_name,initial_name,E_top_list,E_bottom_list,nu_top_list,nu_bottom_list,error_list):
    os.makedirs('./output/{}/{}/Figure/'.format(case_name,initial_name),exist_ok=True)
    print('E_top',E_top_list)
    print('nu_top',nu_top_list)
    print('E_bottom',E_bottom_list)
    print('nu_bottom',nu_bottom_list)
    print('error',error_list)

    plt.figure()
    plt.subplot(1,2,1)
    plt.plot(E_top_list,label=r"$E_{top}$")
    plt.plot(E_bottom_list,label=r"$E_{bottom}$")
    plt.xlabel('Iteration number')
    plt.legend()

    plt.subplot(1,2,2)
    plt.plot(nu_top_list,label=r"$v{top}$")
    plt.plot(nu_bottom_list,label=r"$v{bottom}$")
    plt.xlabel('Iteration number')
    plt.legend()

    plt.tight_layout()
    plt.savefig('./output/{}/{}/Figure/iter.jpg'.format(case_name,initial_name),dpi=2000)

    plt.figure()
    plt.plot(np.log10(error_list))
    plt.title('error')
    plt.ylabel('log(10)')
    plt.savefig('./output/{}/{}/Figure/error.jpg'.format(case_name,initial_name))

    np.savetxt('./output/{}/{}/Figure/E_top_list.txt'.format(case_name,initial_name),E_top_list)
    np.savetxt('./output/{}/{}/Figure/nu_top_list.txt'.format(case_name,initial_name),nu_top_list)
    np.savetxt('./output/{}/{}/Figure/E_bottom_list.txt'.format(case_name,initial_name),E_bottom_list)
    np.savetxt('./output/{}/{}/Figure/nu_bottom_list.txt'.format(case_name,initial_name),nu_bottom_list)
    np.savetxt('./output/{}/{}/Figure/error_list.txt'.format(case_name,initial_name),error_list)
```